\documentclass[runningheads]{llncs}
\usepackage[T1]{fontenc}
\usepackage{graphicx}
\usepackage{amsmath}
\usepackage{amssymb}
\usepackage{textcomp, gensymb}
\usepackage{multirow}
\usepackage[misc,geometry]{ifsym}
\usepackage[colorlinks=true,
            linkcolor=red,
            citecolor=blue]{hyperref}
\usepackage{colortbl,xcolor}
\usepackage{hyperref}
\usepackage[symbol]{footmisc}

\newcommand{\R}{\mathbb{R}}
\newcommand{\C}{\mathbb{C}}
\newcommand{\F}{\mathcal{F}}

\begin{document}
\title{Neural Implicit k-Space for Binning-free Non-Cartesian Cardiac MR Imaging
}

%
%
\author{
Wenqi Huang\inst{1}$^{\textrm{(\Letter)}}$ \and
Hongwei Bran Li\inst{1,2}\and 
Jiazhen Pan\inst{1} \and
Gastao Cruz\inst{3} \and
Daniel~Rueckert\inst{1,4} \and
Kerstin Hammernik\inst{1,4}}
\authorrunning{W. Huang et al.}
%
\institute{
Technical University of Munich, Munich, Germany \and
University of Zurich, Zurich, Switzerland \and
University of Michigan, Michigan, United States \and
Department of Computing, Imperial College London, London, United Kingdom
\\
\email{wenqi.huang@tum.de}
}

\maketitle              
\begin{abstract}
In this work, we propose a novel image reconstruction framework that directly learns a neural implicit representation in k-space for ECG-triggered non-Cartesian Cardiac Magnetic Resonance Imaging (CMR). While existing methods bin acquired data from neighboring time points to reconstruct one phase of the cardiac motion, our framework allows for a \emph{continuous}, \emph{binning-free}, and \emph{subject-specific} k-space representation.
%
%
We assign a unique coordinate that consists of time, coil index, and frequency domain location to each sampled k-space point. 
We then learn the subject-specific mapping from these unique coordinates to k-space intensities using a multi-layer perceptron with frequency domain regularization. During inference, we obtain a complete k-space for Cartesian coordinates and an arbitrary temporal resolution. A simple inverse Fourier transform recovers the image, eliminating the need for density compensation and costly non-uniform Fourier transforms for non-Cartesian data.
This novel imaging framework was tested on 42 radially sampled datasets from 6 subjects. The proposed method outperforms other techniques qualitatively and quantitatively using data from four and one heartbeat(s) and 30 cardiac phases. Our results for one heartbeat reconstruction of 50 cardiac phases show improved artifact removal and spatio-temporal resolution, leveraging the potential for real-time CMR. \footnote[1]{Code available: \url{https://github.com/wenqihuang/NIK_MRI}}

\end{abstract}
\section{Introduction}

Cardiac Magnetic Resonance Imaging (CMR) plays an important role in the clinical assessment of cardiac morphology and function. However, due to the requirement of breath-holds during the data acquisition and the rapid motion of the heart, it is challenging to obtain images with both high temporal and spatial resolution. Considering the limited ability of cardiac patients to perform breath-holds, fast CMR data acquisition with high undersampling rates has attracted great interest in the field of reconstruction.


To reconstruct images from undersampled k-space data, Parallel Imaging (PI) and Compressed Sensing (CS) have been introduced to Magnetic Resonance Imaging (MRI) reconstruction with great success from both the hardware and algorithmic side. 
PI introduced multiple receiver coils to MRI scanners. The coherence among coils enables a higher undersampling rate for data acquisition. 
In CS, sparse priors are widely used in cardiac MRI as regularization terms due to their effective ability to reduce the solution space and avoid local minima of the reconstruction problem. Many of the PI and CS reconstruction algorithms are nowadays available directly on the MRI scanners~\cite{griswold2002grappa,otazo2015low,pruessmann2001advances,pruessmann1999sense,uecker2014espirit}. 


Besides advances in the reconstruction algorithms for MRI, the sampling patterns of MRI have evolved greatly. In clinical practice, Cartesian sampling patterns are commonly used, which can be implemented efficiently on the MRI scanner and facilitate image reconstruction. Here, each data point is equally spaced on a Cartesian grid and can be used directly to perform an inverse fast Fourier transform (FFT) calculation. Although Cartesian sampling is widely used in CMR, it is not robust to motion artifacts. Non-Cartesian coordinate sampling trajectories, such as radial sampling and spiral sampling, were introduced to alleviate this problem. Unlike Cartesian sampling, non-Cartesian sampling can have changeable phase encoding directions during acquisition. Therefore, the noise from moving anatomical structures does not propagate as discrete ghosting along a fixed phase encoding direction but is more distributed over the entire image~\cite{wright2014non}, resulting in more incoherent artifacts. This advantage makes non-Cartesian sampling increasingly popular in CMR. However, the data points are not sampled on a regular grid and therefore require an inverse non-uniform fast Fourier transform (NUFFT) to transform between the k-space and image domains. In NUFFT, the sampled points are re-gridded onto a Cartesian grid using an interpolation kernel.
To compensate for the inhomogeneous distribution of sampled points, a density compensation function (DCF) is additionally required, which is time-consuming and challenging to estimate~\cite{pipe1999sampling,wright2014non}.

Recent developments in deep learning have enabled MRI reconstruction of highly undersampled data. Existing works have shown that exploiting the acquisition physics and the raw k-space data can improve the reconstruction performance substantially~\cite{hammernik2018learning,huang2021lsnet,qin2018convolutional,ramzi2022nc,schlemper2017deep,wang2016accelerating}. These approaches usually require large training databases with fully-sampled k-space data for training. Akçakaya \emph{et al.} proposed a database-free approach to learn k-space interpolation from a fully-sampled k-space center~\cite{akccakaya2019scan}. However, this approach can only be realized for Cartesian sampling patterns. Yaman \emph{et al.} proposed a self-supervised learning approach for Cartesian MRI~\cite{yaman2020self}. Deep image prior was investigated for radially sampled data in~\cite{yoo2021time}. However, this method requires data binning.

As an emerging technique in computer vision, neural implicit functions (NIF) can directly model physical properties from spatial coordinates~\cite{barron2021mip,chibane2020implicit,mildenhall2021nerf}. Such modeling is proven to be effective in shape representation \cite{chibane2020implicit} and scene rendering \cite{mildenhall2021nerf}.  
NIF have attracted increasing attention in medical image analysis tasks, such as radiation therapy \cite{vasudevan2022neural}, super-resolution \cite{vasudevan2022implicit}, shape reconstruction \cite{amiranashvili2022learning}, image registration \cite{wolterink2021implicit}, image segmentation \cite{kuang2022makes} and view reconstruction \cite{zang2021intratomo,zha2022naf}. 
In the context of MRI, Shen \emph{et al.} \cite{shen2022nerp} proposed a sparsely-sampled image reconstruction framework with NIF, which learns a mapping from coordinate to image intensity in the image domain. This approach requires artifact-free images from fully-sampled k-space data for training. In practice, fully-sampled k-space data are hardly available. Additionally, images from undersampled k-space data contain unavoidable global artifacts, which are challenging to remove when NIF operates in the image domain.
To overcome these challenges, we directly learn a mapping \emph{from coordinate to {\mbox{k-space}} signal value} with NIF.

This paper proposes a framework based on neural implicit functions, named neural implicit k-space (NIK), to solve the challenging non-Cartesian CMR reconstruction problem with only a few heartbeats. The proposed NIK framework makes the following three main contributions:
\begin{itemize}
    \item We present a novel k-space point estimation scheme using neural implicit functions for MRI reconstruction, which takes a completely different technical route from the traditional image domain and k-space domain approaches;
    \item We overcome typical challenges in CMR reconstruction: We achieve non-Cartesian reconstruction without NUFFT or conventional gridding, and we can reconstruct an arbitrary number of cardiac phases without loss of temporal information and without data binning.
    \item Our method achieves a single-shot reconstruction, only requiring the imaged object and no additional training data. We compare our method to several single-shot reconstruction methods and it outperforms the baselines with superior qualitative and quantitative results.
\end{itemize}




\section{Methods}
Our proposed method learns a continuous implicit representation of k-space from undersampled non-Cartesian spokes. The representation is optimized with tailored techniques, including frequency domain regularization and high dynamic range loss. After training, the k-space intensities on the Cartesian grid can be generated and converted to the final image by an inverse Fourier transform.
\subsection{Non-Cartesian MR Data Acquisition}



\begin{figure}[!t]
    \centering
    \includegraphics[width=\textwidth]{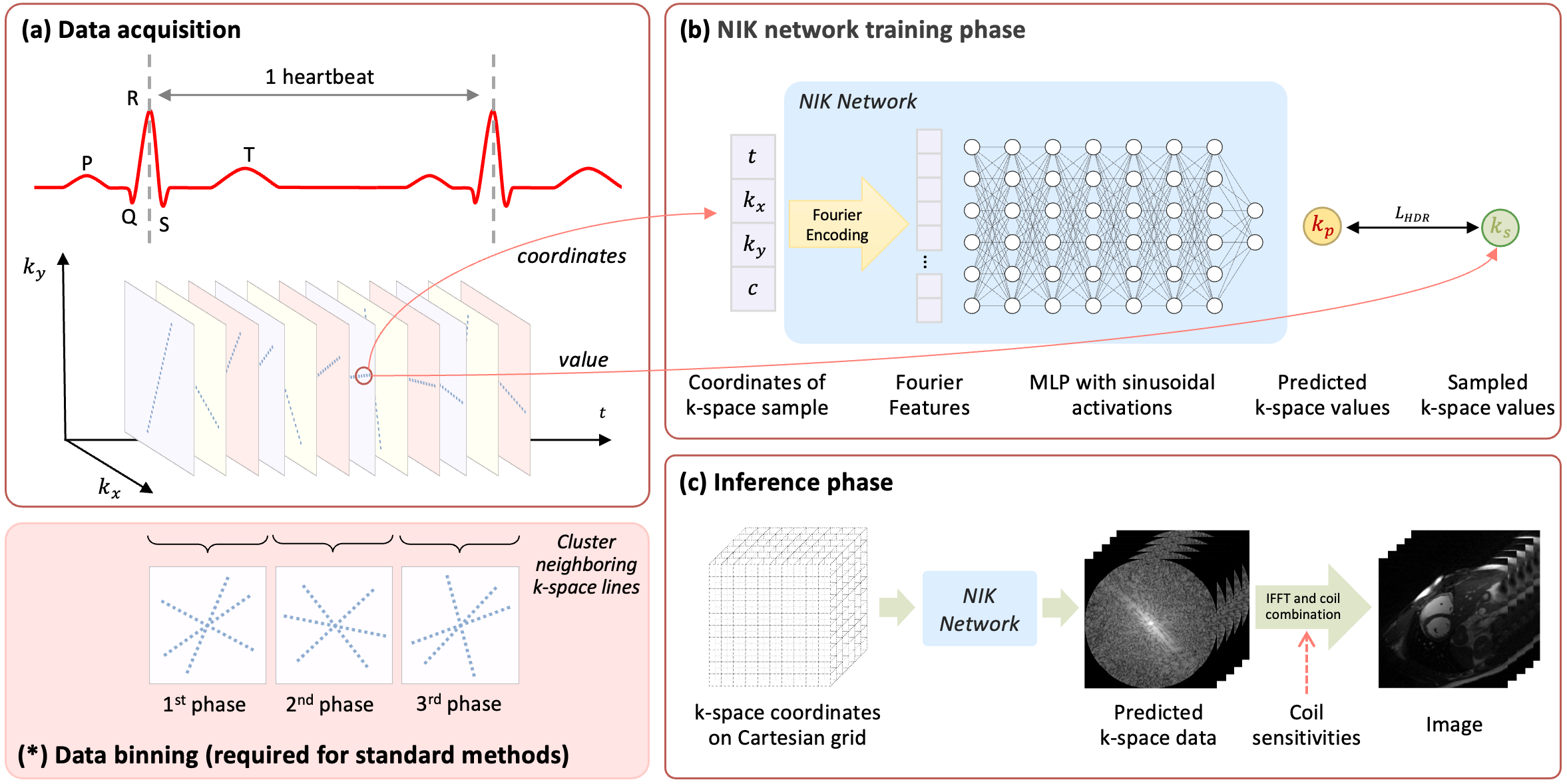}
    \caption{Schematic illustration of neural implicit k-space (NIK). (a) The k-space lines (spokes) are sorted and mapped to one heartbeat. Instead of the traditional data binning (*), we train the MLP to learn the implicit representation of the k-space with the k-space coordinate-intensity pairs (b). $t$, $k_x$, $k_y$, and $c$ refer to time point, k-space coordinates, and coil channel, respectively. (c) In the inference phase, we feed a set of coordinates from the Cartesian grid and obtain the corresponding k-space signal value. The final image can be easily reconstructed by applying the inverse fast Fourier transform and coil combination.}
    \label{fig:framework}
\end{figure}

The signal of an MR image is acquired in a frequency domain, the so-called k-space. Limitations of MR physics, organ motion, and patient compliance allow us to acquire only a limited number of k-space lines (spokes) in a single heartbeat. However, this is often insufficient to reconstruct a high-quality image $x$ with high temporal resolution. To achieve adequate image quality, multiple heartbeat acquisitions during breath-holds and data binning are usually used. While holding the breath virtually eliminates the effect of respiratory motion, capturing the cardiac motion with high temporal resolution remains challenging. For acquisitions across multiple heartbeats, the combined data acquired from different heartbeats need to be sufficiently similar. Otherwise, motion correction is required to account for, e.g., arrhythmia. Fig.~\ref{fig:framework} (a) demonstrates the acquisition and mapping across multiple heartbeats. Time-stamped k-space lines and the ECG signal are acquired, and the ECG signal is used to map all k-space lines to one heartbeat for further reconstruction. Most existing methods divide these k-space lines $y$ into a fixed number of cardiac phases according to their relative timestamps and obtain the binned k-space data $\Tilde{y}$. The neighbouring lines in the time dimension will be clustered together and considered to be the same time point. The reconstruction problem can be formulated as the following standard MR reconstruction problem:
\begin{equation}
    \hat{x} = \arg \min_x \left\|Ax-\Tilde{y}\right\|_2^2 + \lambda R(x),
    \label{eq:cs}
\end{equation}
where $A$ is a non-Cartesian encoding matrix. When only one receiver coil is considered (single coil imaging), $A$ applies NUFFT only to the image $x$. For multi-coil imaging, the different receiver coils are sensitive in different spatial regions as encoded by the coil sensitivity maps. $R(x)$ imposes regularization on $x$ to reduce the solution space and avoid local minima and $\lambda$ is the regularization weight.
However, instead of preserving the original sampling timestamps of the k-space lines, the data binning scheme merges the adjacent k-space lines, assuming they occurred at the same moment, and then reconstructs the cardiac phases at that time point (see Fig.~\ref{fig:framework}(*)). This results in a loss of temporal resolution and lower spatial-temporal coherence exploitation. Notably, data binning will not be required in our proposed method.



\subsection{Neural Implicit k-Space Representation}
For 2D dynamic imaging, the coordinate of each k-space point has four elements: the spatial frequencies $k_x$ and $k_y$, the time $t$, and a coil index $c$. In our setting, the alignment of the data across different heartbeats is still required. However, instead of binning the adjacent k-space lines, we keep their exact timestamps for each point on the k-space lines, which maintains temporal resolution. In this way, each sampled point has its k-space coordinate~$v_i=[t_i, {k_x}_i, {k_y}_i, c_i]^T$ and the corresponding k-space signal value $y_i$, $i=1,2,\dots, N$, and $N$ is the total number of sampled points. Our goal is to model the implicit representation of the continuous k-space by learning a mapping $f:V\rightarrow K$, where $V\subseteq {\R}^4$ is the coordinate space,  and $K\subseteq \C$ is the corresponding complex signal value space. The sampled k-space signal $y_i$ is on a non-Cartesian lattice on $V$. We train a neural network $G_\theta: V\rightarrow K$ with parameters $\theta$ on the data pairs $\left\{(v_i, y_i) | i=1,2,\dots,N\right\}$ to approximate the underlying mapping~$f$. The optimization of parameters $\theta$ is formulated as follows: 
\begin{equation}
    \theta^* = \arg \min_\theta \left\|G_\theta(v) - y\right\|_2^2.
    \label{eq:dc_only}
\end{equation}
Since the coordinate space is a subset of $\R^4$, the coordinates from any kind of sampling trajectory can be used for network training. Importantly, we can query any coordinates during the inference phase. Hence, it is possible to avoid NUFFT by simply feeding the coordinates $\bar{v}$ from a Cartesian grid to the network and obtaining the value of each point on that grid. The reconstructed image $\hat{x}$ can then be obtained by applying the inverse Fourier transform $\F^{-1}$ once:
\begin{equation}
    \hat{x} = \F^{-1}(G_{\theta^*}(\bar{v})).
    \label{eq:inference}
\end{equation}
\subsection{Frequency Domain Regularization}


To avoid local minima and overfitting to the noisy measurements, we introduce the point-wise frequency-domain regularization to Eq. \eqref{eq:dc_only}:
\begin{equation}
    \arg \min_\theta \left\|G_\theta(v) - y\right\|_2^2 + \lambda \left\|G_\theta(v) - D(G_\theta(v)) \right\|_2^2,
    \label{eq:dc_reg}
\end{equation}
where $D$ denotes a frequency-domain denoiser. As the output of the network $G_\theta(v)$ gets closer to the clean signal $y^*$, the denoiser will play a negligible role to ensure $D(G_\theta(v))$ closer to $G_\theta(v)$. However, the choice of $D$ is not trivial. Kernel-based denoising methods have been proposed in the image domain using a set of convolution kernels $g$ to denoise the image $x$. However, these techniques cannot be directly incorporated into our method because the object $G_\theta(v)$ is no longer in the image domain.
We leverage the convolution theorem and take advantage of that the image domain convolution $g*x$ equals the frequency domain multiplication $\F(g)\cdot\F(x)$. Therefore, Eq. \eqref{eq:dc_reg} can be rewritten as:
\begin{equation}
    \arg \min_\theta \left\|G_\theta(v) - y\right\|_2^2 + \lambda \left\|G_\theta(v) - \F(g, v)\cdot G_\theta(v) \right\|_2^2,
    \label{eq:dc_reg2}
\end{equation}
where $\F(g, v)$ denotes the value of position $v$ for the filter kernel $g$ in frequency domain. In this work, we choose the Gaussian kernel as a denoiser kernel. The frequency domain filter value is given as $\F(g, v) = e^{-|v|/2\sigma^2}$, in which $|v|$ is the distance to the k-space center of coordinate $v$ and $\sigma$ is a parameter to control the strength of denoising.

\subsection{High Dynamic Range Loss}
Another optimization challenge in k-space is the large range of k-space values with an imbalanced distribution. The magnitude values around the k-space center which denote the low-frequency components are usually thousands of times higher than the others. If one uses the traditional L2 norm loss in Eq. \eqref{eq:dc_reg2}, the high-magnitude points will be dominant in the training phase and produce an inferior reconstruction result. Inspired by~\cite{mildenhall2022nerf}, we introduce a log transform $\phi(z) = \log(z+\epsilon)$ to the k-space data magnitudes (where $\epsilon$ is a small constant), which compresses the high-magnitude signal to be in comparable magnitude as the low-magnitude ones:
\begin{equation}
    \min_\theta \left\|\phi(G_\theta(v)) - \phi(y)\right\|_2^2 + \lambda \left\|\phi(G_\theta(v)) - \phi(\F(g, v)\cdot G_\theta(v)) \right\|_2^2.
    \label{eq:dc_reg_hdr}
\end{equation}
However, the non-linear transform will also change the noise distribution in the measurement data $y$. To enforce the training to converge to an unbiased result, we approximate the log transform by linearizing it at $G_\theta(v)$, with $\phi'(z) = \frac{1}{z+\epsilon}$:
\begin{equation}
    \min_\theta \left\|\phi'(G_\theta(v))(G_\theta(v) - y)\right\|_2^2 + \lambda \left\|\phi'(G_\theta(v))(G_\theta(v) - \F(g, v)\cdot G_\theta(v)) \right\|_2^2.
    \label{eq:dc_hdr2}
\end{equation}


\subsection{Training and Inference}
During the training phase, we use the acquired data pair $\{(v_i, y_i)| i=1,2,\dots,N\}$ from one single scan to train the network with the loss defined in Eq. \eqref{eq:dc_hdr2}. Once the training is finished, the network has learned the implicit representation of the data distribution in k-space. In inference, we can generate a set of coordinates $\bar{v}$ on the Cartesian grid with an arbitrary temporal resolution. As shown in Eq.~\eqref{eq:inference}, we feed these generated coordinates to the trained network to predict the complex value of these coordinates. Then, a simple inverse Fourier transform is applied to attain the individual coil images. The final image can be obtained by coil combination with coil sensitivity maps. 

\section{Experimental Setup}
\subsubsection{Dataset.}
The proposed NIK framework was evaluated in 6 healthy subjects at a 1.5T scanner (Ingenia, Philips, Best, The Netherlands) with 28 receive coils. Each subject has 7 short-axis slices. To reduce the training time we only used the data from 6 selected coils, discarding coils which are not sensitive in the heart region. All in vivo experiments were conducted with IRB approval and informed consent. The following sequence parameters were used: short axis slice; FOV = $256\times256$ mm$^2$ (considering inherent $1.6\times$ frequency encoding oversampling of radial trajectories);  $8$ mm slice thickness; resolution = $2\times2$ mm$^2$; TE/TR = $1.16/2.3$ ms; b-SSFP readout; radial tiny golden angle of ${\sim}23.6\degree$; flip angle $60\degree$; 8960 radial spokes acquired; nominal scan time ${\sim}20$s; breath-hold acquisition. 

\subsubsection{Coordinate settings.}
The coordinate system for the NIK is flexible and can contain an arbitrary number of continuous or discrete dimensions. In this work, we included only four dimensions: the time dimension $t$, which represents the time point within the averaged heartbeat; the spatial frequencies $k_x$ and $k_y$, which denote the 2D k-space location within the slice plane; the coil index $c$ indicates from which coil the point data originates. All of these coordinates are normalized to $[-1, 1]$ to guarantee their equal contributions to the optimization.

\subsubsection{Model details.}
The proposed NIK framework is based on a multi-layer perceptron (MLP) with Fourier features and periodical activation functions. For each of the input coordinate element $v=[t, {k_x}, {k_y}, c]^T$ we generate a random Gaussian matrix $B$, where each entry is drawn independently from a normal distribution $\mathcal{N}(0, 1)$. The Fourier features position encoding of the coordinate element $v$ is represented as
\begin{equation}
    \gamma(v) = [\cos(2\pi Bv), \sin(2\pi Bv)]
\end{equation}
according to \cite{tancik2020fourier}. The following MLP has 8 layers with a hidden feature dimension of 512. We use Sine activation functions in our MLP according to its effectiveness demonstrated in~\cite{sitzmann2020implicit}. The output layer has two channels, representing the real and imaginary part of the predicted complex k-space value. 

\subsubsection{Training settings.}
During the training, we feed all of the k-space coordinate-value pairs from one or four heartbeats of a single scan to the network. The models were trained by the Adam optimizer with parameters $\beta_1 = 0.9$, $\beta_2 = 0.999$, $\varepsilon = 10^{-8}$, learning rate $3\cdot 10^{-5}$, and batch size 1. The training procedure stopped after 50000 steps when the loss did not decrease any further. The models were implemented with PyTorch 1.12. The training and testing were performed on an NVIDIA A6000 GPU. The regularization weight $\lambda$ was empirically selected as 0.4 and 0.5 for four heartbeats and one heartbeat reconstruction, respectively. 
The $\epsilon$ in training loss was set to $0.01$, and $\sigma=1$ for frequency-domain regularization.
The network training took approximately 10 hours with four heartbeats data, and 3 hours for one heartbeat. The reconstruction of 30 cardiac phases with a resolution of $256\times 256$ after training is less than 30 seconds.


\subsubsection{Comparison of methods.}
Data were reconstructed for 30 cardiac phases using inverse NUFFT (INUFFT), Conjugate Gradient SENSE (CG-SENSE)~\cite{pruessmann2001advances}, Low Rank plus Sparse (L+S)~\cite{otazo2015low}, and the proposed NIK with one heartbeat and four heartbeats. For comparison methods, there are about 14 and 55 spokes for each cardiac phase for one heartbeat and four heartbeats, respectively. Reference CG-SENSE reconstructions with twenty heartbeats (${\sim}$270 spokes per cardiac phase) were considered to evaluate the performance of all methods.


\subsubsection{Evaluation metrics.}
Both visual comparison and quantitative evaluation were used for performance evaluation. For a quantitative evaluation, the normalized root mean squared error (NRMSE), peak signal-to-noise ratio (PSNR) and structural similarity index (SSIM) were calculated. A higher PSNR and SSIM and lower NRMSE indicate better quantitative performance.


\begin{table}[!t]
  \caption{\label{singlecoiltab}The average MSE, PSNR and SSIM of INUFFT, CG-SENSE, L+S and the proposed NIK for 4 heartbeats and 1 heartbeat (mean±std).}
  \centering

  \begin{tabular}{ccccc}
      \hline\hline
      Heartbeat & Methods & NRMSE	& PSNR (dB) & SSIM	\\
      \hline
      \multirow{4}{1em}{4} 
          &INUFFT	&0.38±0.08	&28.18±2.07	&0.58±0.08\\
          &CG-SENSE \cite{pruessmann2001advances}	&0.23±0.04	&31.38±1.84	&0.75±0.05\\
          &L+S \cite{otazo2015low} &0.22±0.04 & 32.06±1.69 &0.76±0.04\\
          &{\bf NIK (proposed)}	&{\bf 0.19±0.04}	&{\bf 33.52±2.08}	&{\bf 0.84±0.05}\\
    
      \hline
      \multirow{4}{1em}{1} 
          &INUFFT	&0.79±0.14	&21.53±1.52	&0.34±0.06\\
          &CG-SENSE \cite{pruessmann2001advances}&0.43±0.06	&25.48±1.28	&0.54±0.06\\
          &L+S \cite{otazo2015low}&0.35±0.04 & 27.77±1.44 &0.65±0.04\\
          &{\bf NIK (proposed)}	&{\bf 0.25±0.05}	&{\bf 30.67±1.94}	&{\bf 0.77±0.05}\\

    
    
      \hline\hline
  \end{tabular}
    \label{tab:metrics}
\end{table}

\section{Results}

\begin{figure}
    \centering
    \includegraphics[width=0.95\textwidth]{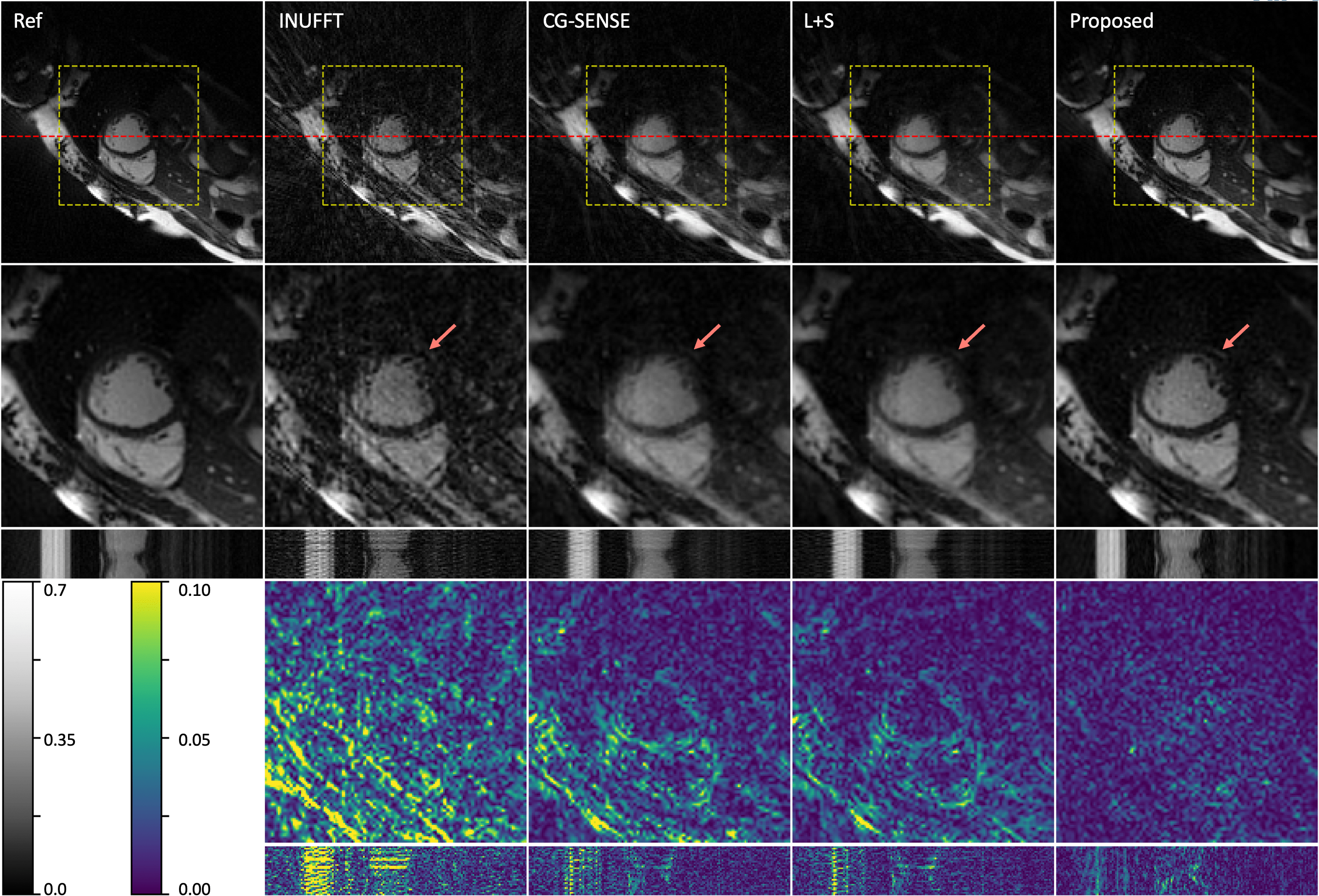}
    \caption{Reconstructions with data from four heartbeats. The 2$^\text{nd}$ and 4$^\text{th}$ rows show the zoomed views of the yellow boxes and their error maps. The border of the myocardium is marked by red arrows. The 3$^\text{rd}$ and 5$^\text{th}$ row show the x-t profile and their error maps.}
    \label{fig:4hb}
\end{figure}

\begin{figure}
    \centering
    \includegraphics[width=0.95\textwidth]{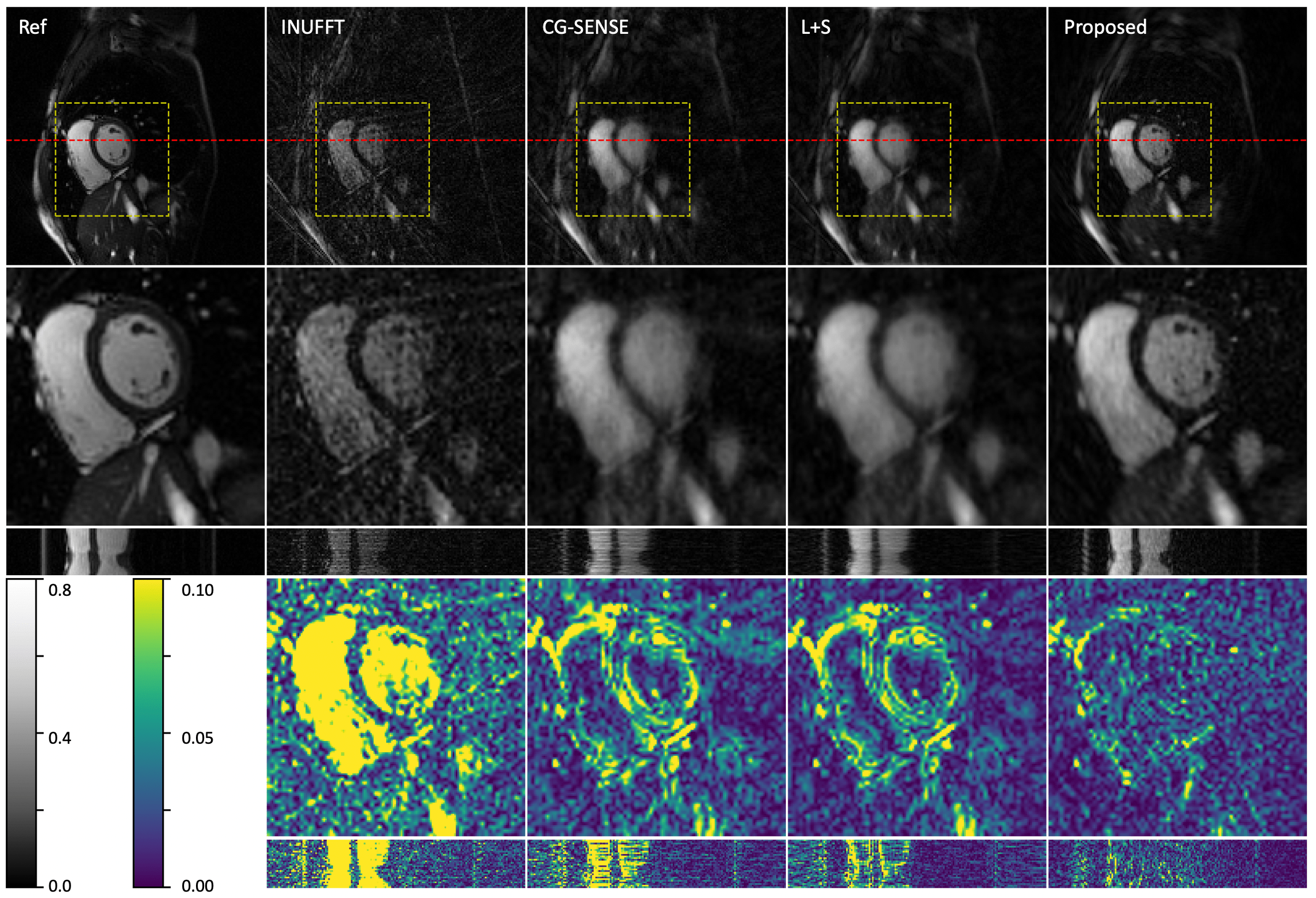}
    \caption{Reconstructions with data from one heartbeat. The 2$^\text{nd}$ and 4$^\text{th}$ rows show the zoomed views of the yellow boxes and their error maps. The 3$^\text{rd}$ and 5$^\text{th}$ rows show the x-t profile and their error maps.}
    \label{fig:1hb}
\end{figure}

Table~\ref{tab:metrics} summarized the mean and standard deviation of the reconstruction results for 42 slices from 6 subjects. For the four heartbeats and one heartbeat reconstructions, we can see that our proposed NIK consistently outperforms the comparison methods. A visual comparison of the four heartbeat reconstruction can be found in Fig.~\ref{fig:4hb}, which supports our quantitative findings. From the zoomed images in the second row, we notice that the comparison methods suffer from the loss of small anatomical structures (red arrows), while NIK is able to recover them. x-t profiles show larger errors for the comparison methods with respect to the temporal dimension. This is related to the striking artifacts produced by radial sampling in the 2D plane, which are challenging to eliminate if the conventional NUFFT is used. In contrast, our proposed method reconstructs the temporal information coherently and accurately. The reconstruction of one heartbeat is showcased in Fig.~\ref{fig:1hb}. Due to the extremely small number of spokes, INUFFT, CG-SENSE and L+S end up either with noisy images or blurred images. 
In contrast, NIK is competent in recovering sharp images with high temporal resolution.

\begin{figure}[t!]
    \centering
    \includegraphics[width=\textwidth]{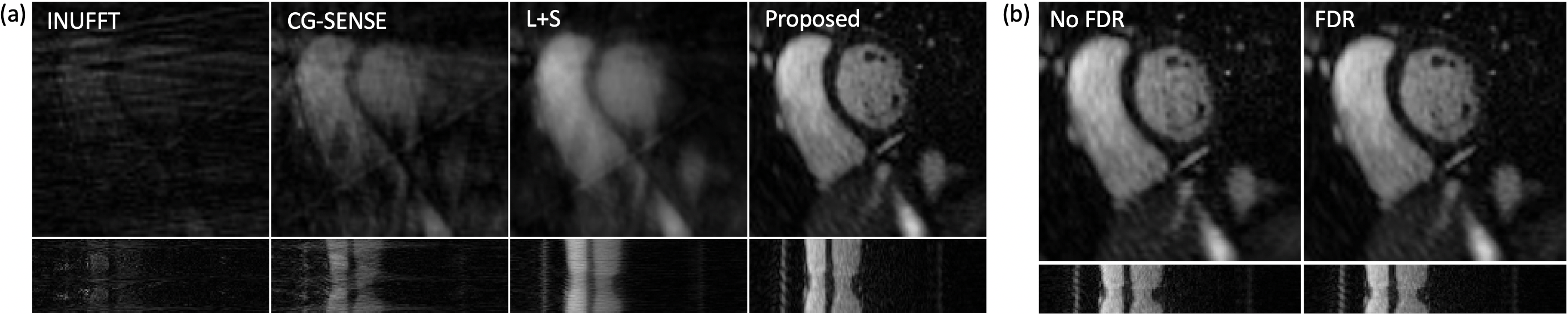}
    \caption{(a) Reconstructions for 50 cardiac phases  and (b) influence of frequency-domain regularization (FDR) with 30 cardiac phases, with data from one heartbeat.}
    \label{fig:50f_and_reg}
\end{figure}

Fig.~\ref{fig:50f_and_reg}(a) shows the results for 50 cardiac phases of one heartbeat reconstruction. The reconstruction performance of INUFFT, CG-SENSE and L+S is much inferior compared to Fig.~\ref{fig:1hb} and the images are overwhelmed by the artifacts caused by radial sampling. Because of the data-binning scheme, the number of spokes available for each cardiac phase dropped to 3/5 of the original when we increase from 30 to 50 phases ($\sim$8 spokes per cardiac phase). While the proposed NIK still has similar image quality as 30 phase reconstruction.
We also demonstrated the effectiveness of frequency domain regularization in Fig.~\ref{fig:50f_and_reg}(b). The result with frequency domain regularization shows slightly reduced noise.



\section{Discussion and Conclusion}
In this work, we present the concept of neural implicit k-space (NIK) for non-Cartesian CMR. NIK is a new database-free learning framework to resolve temporal information in a continuous non-Cartesian acquisition paradigm. Instead of merging adjacent k-space lines in the temporal dimension to reconstruct each cardiac phase, a continuous implicit representation of k-space can be learned so that k-space points on a Cartesian grid can be generated and transformed into clean images by inverse Fourier transform. In this paper, we focus on radially sampled CMR data. We believe that NIK can be easily extended to arbitrary sampling patterns and N-dimensional datasets by modifying the input coordinate. More sampling patterns and modalities will be investigated.

Our experiments illustrate that NIK can reconstruct images in high quality and shows superior performance in removing artifacts caused by radial sampling patterns. NIK outperforms all reference methods by a large margin qualitatively and quantitatively. 
As NIK learns a continuous k-space representation and is binning-free, the image quality is not affected by the number of cardiac phases to be reconstructed. This is not possible for conventional methods based on data binning, as some phases might have too few k-space lines.
The results for one heartbeat reconstruction show promise for real-time cardiac imaging, even though some small details are lost as they might not be encoded in the acquired k-space. Noise can still remain in the image reconstructions, which is well-known from Cartesian k-space interpolation techniques~\cite{akccakaya2019scan,griswold2002grappa}. This suggests future research in frequency-domain filtering beyond Gaussian filtering. 

 
Our method achieves a single-shot reconstruction that does not require a training dataset from a large population, as it is commonly demanded by deep learning-based reconstruction methods. While enabling single heartbeat reconstruction, we acknowledge that the long training time (i.e., several hours) is the main limitation of our approach. 
This is caused by the slow converge during optimization given the large range nature of sampled k-space values. In our future work, we will consider pre-training techniques to shorten the training time and improve the reconstruction speed.
%


In summary, we introduce a novel concept of neural implicit k-space (NIK) for database-free k-space interpolation of non-Cartesian CMR data. The k-space pointwise estimation scheme comes naturally with the advantages of binning-free, sampling pattern-independent, and NUFFT-free imaging. This work demonstrates great potential to leverage NIK in further CMR applications and beyond. 

\subsubsection{Acknowledgements} This work was supported in part by the European Research Council (Grant Agreement no. 884622).

%
%
%
\bibliographystyle{splncs04}
\bibliography{ref}

\begin{thebibliography}{10}
\providecommand{\url}[1]{\texttt{#1}}
\providecommand{\urlprefix}{URL }
\providecommand{\doi}[1]{https://doi.org/#1}

\bibitem{akccakaya2019scan}
Ak{\c{c}}akaya, M., Moeller, S., Weing{\"a}rtner, S., U{\u{g}}urbil, K.:
  {Scan-specific robust artificial-neural-networks for k-space interpolation
  (RAKI) reconstruction: Database-free deep learning for fast imaging}.
  Magnetic resonance in medicine  \textbf{81}(1),  439--453 (2019)

\bibitem{amiranashvili2022learning}
Amiranashvili, T., L{\"u}dke, D., Li, H.B., Menze, B., Zachow, S.: Learning
  shape reconstruction from sparse measurements with neural implicit functions.
  In: International Conference on Medical Imaging with Deep Learning. pp.
  22--34. PMLR (2022)

\bibitem{barron2021mip}
Barron, J.T., Mildenhall, B., Tancik, M., Hedman, P., Martin-Brualla, R.,
  Srinivasan, P.P.: {Mip-NeRF: A multiscale representation for anti-aliasing
  neural radiance fields}. In: Proceedings of the IEEE/CVF International
  Conference on Computer Vision. pp. 5855--5864 (2021)

\bibitem{chibane2020implicit}
Chibane, J., Alldieck, T., Pons-Moll, G.: {Implicit functions in feature space
  for 3D shape reconstruction and completion}. In: Proceedings of the IEEE/CVF
  Conference on Computer Vision and Pattern Recognition. pp. 6970--6981 (2020)

\bibitem{griswold2002grappa}
Griswold, M.A., Jakob, P.M., Heidemann, R.M., Nittka, M., Jellus, V., Wang, J.,
  Kiefer, B., Haase, A.: {Generalized autocalibrating partially parallel
  acquisitions (GRAPPA)}. Magnetic Resonance in Medicine: An Official Journal
  of the International Society for Magnetic Resonance in Medicine
  \textbf{47}(6),  1202--1210 (2002)

\bibitem{hammernik2018learning}
Hammernik, K., Klatzer, T., Kobler, E., Recht, M.P., Sodickson, D.K., Pock, T.,
  Knoll, F.: {Learning a variational network for reconstruction of accelerated
  MRI data}. Magnetic resonance in medicine  \textbf{79}(6),  3055--3071 (2018)

\bibitem{huang2021lsnet}
Huang, W., Ke, Z., Cui, Z.X., Cheng, J., Qiu, Z., Jia, S., Ying, L., Zhu, Y.,
  Liang, D.: {Deep low-rank plus sparse network for dynamic MR imaging}.
  Medical Image Analysis  \textbf{73},  102190 (2021)

\bibitem{kuang2022makes}
Kuang, K., Zhang, L., Li, J., Li, H., Chen, J., Du, B., Yang, J.: What makes
  for automatic reconstruction of pulmonary segments. In: Medical Image
  Computing and Computer Assisted Intervention--MICCAI 2022: 25th International
  Conference, Singapore, September 18--22, 2022, Proceedings, Part I. pp.
  495--505. Springer (2022)

\bibitem{mildenhall2022nerf}
Mildenhall, B., Hedman, P., Martin-Brualla, R., Srinivasan, P.P., Barron, J.T.:
  {NeRF in the dark: High dynamic range view synthesis from noisy raw images}.
  In: Proceedings of the IEEE/CVF Conference on Computer Vision and Pattern
  Recognition. pp. 16190--16199 (2022)

\bibitem{mildenhall2021nerf}
Mildenhall, B., Srinivasan, P.P., Tancik, M., Barron, J.T., Ramamoorthi, R.,
  Ng, R.: {NeRF: Representing scenes as neural radiance fields for view
  synthesis}. Communications of the ACM  \textbf{65}(1),  99--106 (2021)

\bibitem{otazo2015low}
Otazo, R., Candes, E., Sodickson, D.K.: {Low-rank plus sparse matrix
  decomposition for accelerated dynamic MRI with separation of background and
  dynamic components}. Magnetic resonance in medicine  \textbf{73}(3),
  1125--1136 (2015)

\bibitem{pipe1999sampling}
Pipe, J.G., Menon, P.: {Sampling density compensation in MRI: rationale and an
  iterative numerical solution}. Magnetic Resonance in Medicine: An Official
  Journal of the International Society for Magnetic Resonance in Medicine
  \textbf{41}(1),  179--186 (1999)

\bibitem{pruessmann2001advances}
Pruessmann, K.P., Weiger, M., B{\"o}rnert, P., Boesiger, P.: {Advances in
  sensitivity encoding with arbitrary k-space trajectories}. Magnetic Resonance
  in Medicine: An Official Journal of the International Society for Magnetic
  Resonance in Medicine  \textbf{46}(4),  638--651 (2001)

\bibitem{pruessmann1999sense}
Pruessmann, K.P., Weiger, M., Scheidegger, M.B., Boesiger, P.: {SENSE:
  sensitivity encoding for fast MRI}. Magnetic Resonance in Medicine: An
  Official Journal of the International Society for Magnetic Resonance in
  Medicine  \textbf{42}(5),  952--962 (1999)

\bibitem{qin2018convolutional}
Qin, C., Schlemper, J., Caballero, J., Price, A.N., Hajnal, J.V., Rueckert, D.:
  {Convolutional recurrent neural networks for dynamic MR image
  reconstruction}. IEEE transactions on medical imaging  \textbf{38}(1),
  280--290 (2018)

\bibitem{ramzi2022nc}
Ramzi, Z., Chaithya, G., Starck, J.L., Ciuciu, P.: {NC-PDNet: a
  density-compensated unrolled network for 2D and 3D non-Cartesian MRI
  reconstruction}. IEEE Transactions on Medical Imaging  (2022)

\bibitem{schlemper2017deep}
Schlemper, J., Caballero, J., Hajnal, J.V., Price, A., Rueckert, D.: {A deep
  cascade of convolutional neural networks for MR image reconstruction}. In:
  International conference on information processing in medical imaging. pp.
  647--658. Springer (2017)

\bibitem{shen2022nerp}
Shen, L., Pauly, J., Xing, L.: {NeRP: implicit neural representation learning
  with prior embedding for sparsely sampled image reconstruction}. IEEE
  Transactions on Neural Networks and Learning Systems  (2022)

\bibitem{sitzmann2020implicit}
Sitzmann, V., Martel, J., Bergman, A., Lindell, D., Wetzstein, G.: {Implicit
  neural representations with periodic activation functions}. Advances in
  Neural Information Processing Systems  \textbf{33},  7462--7473 (2020)

\bibitem{tancik2020fourier}
Tancik, M., Srinivasan, P., Mildenhall, B., Fridovich-Keil, S., Raghavan, N.,
  Singhal, U., Ramamoorthi, R., Barron, J., Ng, R.: {Fourier features let
  networks learn high frequency functions in low dimensional domains}. Advances
  in Neural Information Processing Systems  \textbf{33},  7537--7547 (2020)

\bibitem{uecker2014espirit}
Uecker, M., Lai, P., Murphy, M.J., Virtue, P., Elad, M., Pauly, J.M.,
  Vasanawala, S.S., Lustig, M.: {ESPIRiT—an eigenvalue approach to
  autocalibrating parallel MRI: where SENSE meets GRAPPA}. Magnetic resonance
  in medicine  \textbf{71}(3),  990--1001 (2014)

\bibitem{vasudevan2022neural}
Vasudevan, V., Shen, L., Huang, C., Chuang, C., Islam, M., Ren, H., Yang, Y.,
  Dong, P., Xing, L.: {Neural representation for three-dimensional dose
  distribution and its applications in precision radiation therapy}.
  International Journal of Radiation Oncology, Biology, Physics
  \textbf{114}(3), ~e552 (2022)

\bibitem{vasudevan2022implicit}
Vasudevan, V., Shen, L., Huang, C., Chuang, C., Islam, M.T., Ren, H., Yang, Y.,
  Dong, P., Xing, L.: {Implicit neural representation for radiation therapy
  dose distribution}. Physics in Medicine \& Biology  \textbf{67}(12),  125014
  (2022)

\bibitem{wang2016accelerating}
Wang, S., Su, Z., Ying, L., Peng, X., Zhu, S., Liang, F., Feng, D., Liang, D.:
  {Accelerating magnetic resonance imaging via deep learning}. In: 2016 IEEE
  13th international symposium on biomedical imaging (ISBI). pp. 514--517. IEEE
  (2016)

\bibitem{wolterink2021implicit}
Wolterink, J.M., Zwienenberg, J.C., Brune, C.: {Implicit neural representations
  for deformable image registration}. In: Medical Imaging with Deep Learning
  (2021)

\bibitem{wright2014non}
Wright, K.L., Hamilton, J.I., Griswold, M.A., Gulani, V., Seiberlich, N.:
  {Non-Cartesian parallel imaging reconstruction}. Journal of Magnetic
  Resonance Imaging  \textbf{40}(5),  1022--1040 (2014)

\bibitem{yaman2020self}
Yaman, B., Hosseini, S.A.H., Moeller, S., Ellermann, J., U{\u{g}}urbil, K.,
  Ak{\c{c}}akaya, M.: {Self-supervised learning of physics-guided
  reconstruction neural networks without fully sampled reference data}.
  Magnetic resonance in medicine  \textbf{84}(6),  3172--3191 (2020)

\bibitem{yoo2021time}
Yoo, J., Jin, K.H., Gupta, H., Yerly, J., Stuber, M., Unser, M.:
  {Time-dependent deep image prior for dynamic MRI}. IEEE Transactions on
  Medical Imaging  \textbf{40}(12),  3337--3348 (2021)

\bibitem{zang2021intratomo}
Zang, G., Idoughi, R., Li, R., Wonka, P., Heidrich, W.: {IntraTomo:
  self-supervised learning-based tomography via sinogram synthesis and
  prediction}. In: Proceedings of the IEEE/CVF International Conference on
  Computer Vision. pp. 1960--1970 (2021)

\bibitem{zha2022naf}
Zha, R., Zhang, Y., Li, H.: {NAF: neural attenuation fields for sparse-view
  CBCT Reconstruction}. In: International Conference on Medical Image Computing
  and Computer-Assisted Intervention. pp. 442--452. Springer (2022)

\end{thebibliography}
%




\end{document}